\documentclass[aps,pre,twocolumn,amsmath,amssymb,superscriptaddress,showpacs,floatfix]{revtex4}
\usepackage{graphicx}

\begin{document}

\title{Diffusion in the presence of scale-free absorbing boundaries}
\author{Nir Alfasi}
\address{Raymond and Beverly Sackler School of Physics and
Astronomy, Tel Aviv University, Tel Aviv 69978, Israel}
\address{Department of Electrical Engineering, Technion - Israel Institute of Technology, Haifa 32000, Israel}
\author{Yacov Kantor}
\address{Raymond and Beverly Sackler School of Physics and
Astronomy, Tel Aviv University, Tel Aviv 69978, Israel}
\date{\today}

\begin{abstract}
Scale-free surfaces, such as cones, remain unchanged under
a simultaneous expansion of all coordinates by the same factor.
Probability density of a particle diffusing near such absorbing
surface at large time approaches a simple form that incorporates
power-law dependencies on time and distance from a special point,
such as apex of the cone, which are  characterized by a single
exponent $\eta$. The same exponent is used to describe the number
of spatial conformations of long ideal polymer attached to the
special point of a repulsive surface of the same geometry and can be
used in calculation of entropic forces between such polymers and
surfaces. We use the solution of diffusion equation near such
surfaces to find the  numerical values of $\eta$, as well as
to provide some insight into the behavior of ideal polymers near
such surfaces.

\end{abstract}
\pacs{
05.40.-a 
64.60.F- 
82.35.Lr 
02.60.Cb 
 }
\maketitle

\section{Introduction}
\begin{figure}[t]
	\centerline{\includegraphics[width=8.5cm]{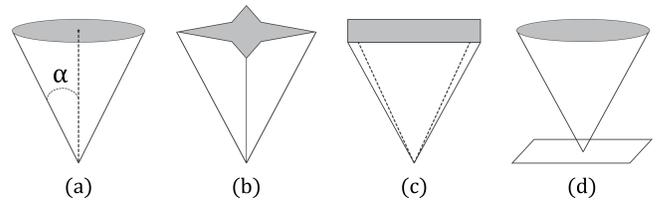}}
\caption{Examples of scale-free surfaces: cones with (a)
circular, (b) star-shaped and
(c) square cross-sections, and (d)
a circular cone touching a plane. All shapes are infinite, with gray
surfaces indicating truncation for graphic purposes.}
\label{fig:scalefree}
\end{figure}

Diffusion in the presence of absorbing boundaries is a well
explored problem \cite{Carslaw1959,Redner2001}. It remains
an active field of research due to its importance to numerous fields
in physics, chemistry, biology and economy
\cite{Weiss1994,Berg1993,Kimura1964,Abraham2002}. In particular, it
is related to the problem of first-passage processes
\cite{Hughes1995,Redner2001,Ben-Naim2010,Ben-Naim14}. In this work,
we consider the diffusion of a particle near {\em scale-free}  (SF)
surfaces, such as cones of different cross-sections
[Figs.~\ref{fig:scalefree}(a)-\ref{fig:scalefree}(c)] or a
combination of a cone and a
plane [Fig.~\ref{fig:scalefree}(d)]. (Additional examples of SF
shapes can be seen in Fig. 1 in Ref.~\cite{Hammer2014}.) Such
surfaces have no  characteristic length scale, i.e., their {\em shape}
is not modified under rescaling by an arbitrary factor $\lambda$; i.e.,
$\vec{r}\to\lambda\vec{r}$.  Generally, such scale transformation
changes the position of the surface. We will always choose the
origin of coordinates as special point on the surfaces, such as apex
of the cone, ensuring that the position does not change either.
This point will also play an important role in the physical problem:
in diffusion problem the particle will be released in the neighborhood
of that point, while in the polymer problem one end of the polymer
will be held in that vicinity.

The absence of a geometric length scale leads to a rather interesting
behavior of the solutions of the diffusion problem, as we explain
in Sec.~\ref{sec:diffusion}. In general, the
diffusion problem can be related to statistical mechanics of {\em ideal
polymers} in which self-interactions can be neglected
\cite{Gennes1979,Rubinstein2003,Doi1996}. In the case of SF surfaces
the solutions of diffusion equations can be used to infer the prefactor
in force-distance relation characterizing interactions between ideal
polymers and the surfaces \cite{Maghrebi2011,Maghrebi2012,Hammer2014}.
Section \ref{sec:polymer} explains the relation between the diffusion
and ideal polymer problems, as well as discusses some general features
of polymers near surfaces. In this work we employ a simple numerical
approach to the problem and  demonstrate its usefulness to the
quantitative solution of  polymer-surface interaction. In Sec.~\ref{sec:method} we describe our numerical approach, and in Sec.~\ref{sec:results} we use such solutions to gain intuitive
insights into the behavior of ideal polymers near repulsive surfaces
that have no azimuthal symmetry.

\section{Diffusion near scale-free surfaces}\label{sec:diffusion}
In its most elementary form the diffusion process is described by the
probability density $P(\vec{r},\vec{r}_0,t)$ of finding a diffusing
particle at a position $\vec{r}$ at time $t$, if at $t=0$ it was
located at $\vec{r}_0$. Such probability satisfies the diffusion equation
\begin{equation}
\frac{\partial P}{\partial t}=D\nabla^2 P,
\label{eq:diffeq}
\end{equation}
which must be supplemented by the initial condition
$P(\vec{r},\vec{r}_0,t=0)=\delta(\vec{r}-\vec{r}_0)$, as well as by
the boundary conditions. In the presence of {\em absorbing} boundaries
it is required that  $P$ vanishes when $\vec{r}$ is on the boundary.
Equation \eqref{eq:diffeq} corresponds to diffuser in continuous space
performing infinitesimal steps. For a random walker on a
$d$-dimensional periodic (e.g., hypercubic) lattice with lattice
constant $a$, with dimensionless time measuring the number of discrete
steps, the Laplacian in Eq.~\eqref{eq:diffeq} is replaced by its
discrete version, while the time derivative of $P$ becomes a difference
in probabilities at (dimensionless) times $t+1$ and $t$. In this case
the diffusion constant becomes $D=a^2/2d$. (The theory presented here
is valid at arbitrary $d$, but the numerical examples will focus on
simulations performed on a three-dimensional cubic lattice.)
In the presence of absorbing boundaries, the survival probability
$S(\vec{r}_0,t)=\int P(\vec{r},\vec{r}_0,t){\rm d}^dr$
of a particle that at time $t=0$ was at point $\vec{r}_0$ decays with
time. The function $S$ also satisfies \cite{Wiegel86} diffusion Eq.~\eqref{eq:diffeq} with spatial derivatives taken with respect to variable
$\vec{r}_0$, but its initial condition is $S(\vec{r}_0,t=0)=1$ inside the
permitted volume while vanishing (at any $t$) on the absorbing boundary.

The trace of a particle diffusing on a lattice for time $t=N$ can be
viewed as a configuration of an ideal polymeric chain with $N+1$
monomers \cite{Gennes1979,Rubinstein2003,Doi1996}. In the absence of
confining surfaces the number of different configurations starting at
a specific lattice site $\vec{r}_0$ is ${\cal N}(\vec{r}_0,N)=z^N$, where
$z$ is the coordination number of a lattice. If {\em repulsive} boundaries
are present, the configurations that cross the boundary must be excluded. This is
accounted for by considering a random walk (RW) problem with {\em absorbing}
boundary conditions, and ${\cal N}(\vec{r}_0,N)=z^NS(\vec{r}_0,N)$.
Thus solution of the diffusion problem provides a handle on counting
the configurations of ideal polymers. The latter determines the free
energy of the polymers and can be used to find forces between the
polymers and confining surfaces \cite{Maghrebi2011,Maghrebi2012}.

Solution of diffusion Eq.~\eqref{eq:diffeq} in a {\em finite}
space surrounded by absorbing surfaces can be presented in the
form \cite{Morse53}
$P(\vec{r},\vec{r}_0,t)=\sum_iA_i\Phi_i(\vec{r}){\rm e}^{-t/\tau_i}$,
where $\Phi_i$ and $1/\tau_i$ are the eigenfunctions and eigenvalues
of the equation $\nabla^2\Phi_i=-\Phi_i/\tau_i$, while the prefactors
$A_i$ depend on the initial position $\vec{r}_0$ of the particle.
Similarly, the survival probability
$S(\vec{r}_0,t)=\sum_i B_i\Phi_i(\vec{r}_0) {\rm e}^{-t/\tau_i}$
\cite{Morse53}. For long times the behavior of both functions will
be dominated by the smallest eigenvalue
$1/\tau_1$, i.e., $P\propto \Phi_1(\vec{r}){\rm e}^{-t/\tau_1}$, and
$S\propto \Phi_1(\vec{r}_0){\rm e}^{-t/\tau_1}$. Typically, the value
of the largest time $\tau_1\sim \ell^2/D$, where $\ell$ is the largest
linear dimension of the confining volume. If the confining volume is
{\em not finite} the solution never reaches the state of exponential
decay. In this work, we consider the diffusion of a particle near SF
surfaces. Since both the equation, and the boundary and initial
conditions imposed on $S(\vec{r}_0,t)$ are SF, $S$ can only depend on
the scaled (dimensionless) variable $\vec{w}=\vec{r}_0/\sqrt{Dt}$.
Moreover, for $w\ll 1$ the Eq.~\eqref{eq:diffeq} reduces to
$\nabla_{\vec{w}}^2S=0$, where the derivatives of the Laplacian are
taken with respect to $\vec{w}$. In this limit we expect a simple
power law solution \cite{Ben-Naim2010}
\begin{equation}
S= w^\eta g(\{\theta_i\})=(r_0/\sqrt{Dt})^\eta g(\{\theta_i\}),
\label{eq:S_eta}
\end{equation}
where $g$ is a function of $d-1$ angular variables $\{\theta_i\}$
\cite{Hammer2014}, which solves the eigenvalue equation
\begin{equation}
\Delta_{S^{d-1}}g=\eta(2-d-\eta)g,
\label{eq:eta}
\end{equation}
where $\Delta_{S^{d-1}}$ is the spherical Laplacian \cite{Rosenberg1997}.
Angular variables represent position on a unit sphere and we expect a
discrete spectrum of values of $\eta=\eta_i$. We are interested
in large $t$ (small $w$) limit, and therefore choose $\eta$ to be
the smallest eigenvalue of Eq.~\eqref{eq:eta}.  Since the probability
density must be positive, the eigenfunction corresponding to that $\eta$
is a function that does not change sign \cite{Ben-Naim2010}. We note that
Eqs.~\eqref{eq:S_eta} and \eqref{eq:eta}, as well as the equations
that will be derived from them in Sec. \ref{sec:method}, are valid for
any SF surfaces and do not require additional symmetries, such as
azimuthal symmetry.

\section{Diffusion as a polymer problem}\label{sec:polymer}

Many properties of polymers in free space \cite{Gennes1979} and
near limiting  surfaces \cite{Binder83,debell,Eisenriegler1993}
can be deduced from the theory of critical phenomena. In particular,
some behaviors are characterized by critical exponents, that
are independent of microscopic details of the Hamiltonian.  Such universality
enables usage of simplified models, such as RWs on lattices representing
ideal polymers, or self-avoiding walks (SAWs) representing polymers in
good solvent \cite{Gennes1979,remark}.  On a regular lattice, the number of
conformations of  a polymer in free space is $\mathcal{N}\propto z^N N^{\gamma-1}$,
where $z$ is the lattice coordination number in the case of a RW,
and an effective coordination number in the case of SAW. Critical
exponent $\gamma$  is universal \cite{Gennes1979}. This exponent
is related by Fisher's identity \cite{Cardy1996} to correlation
length exponent $\nu$ and the exponent $\eta$ characterizing the
anomalous decay of density correlations: $\gamma=(2-\eta)\nu$.
Values of the exponents $\gamma$, $\nu$ and $\eta$ differ between
RWs and SAWs. In this paper we deal only with {\em ideal polymers}
model for which $\nu=1/2$ \cite{remark}.

Presence of scale-invariant boundaries modifies the behavior of
polymers. In the expression for the number of configurations
$\mathcal{N}\propto z^N N^{\gamma-1}$, the leading {\em non-universal}
part $z^N$ remains unchanged! Similarly, there is no change
in the correlation length exponent
$\nu$ \cite{Binder83,debell,Eisenriegler1993}. However, the
exponent $\gamma$  characterizing the {\em subleading}
$N$-dependence of $\mathcal{N}$ changes its value. (For flat surfaces
it is frequently denoted $\gamma_1$ \cite{Binder83}, while for wedges
with opening angle $\alpha$ it is sometimes denoted $\gamma_2(\alpha)$
\cite{Cardy_Redner84}.) As their free space counterparts, these exponents
do not depend on the microscopic details of the Hamiltonian, but their
values {\em do depend} on the type of the
limiting surface. Exponent $\eta$ describing the behavior of the
correlation functions is also affected by the presence of the surface.
However, Fisher's relation, which is a consequence of the fact that
the total number of states is an integral of the correlation function,
remains valid even in those modified circumstances; e.g., for a flat
surface $\gamma_1=(2-\eta_\perp)\nu$, where $\eta_\perp$ describes
the power-law dependence of the correlation function in any direction
except parallel to the surface \cite{Cardy_Redner84}. This relation
persists also for wedges and cones. In our problem, all the exponents
correspond to specific surfaces under consideration and we omit the
various subscripts used in the literature. Since we are considering only
{\em ideal polymers}, Fisher's relation for our surface-specific exponents
reduces to $\gamma-1=-\eta/2$ leading to  $\mathcal{N}\propto z^N N^{-\eta/2}$.
By comparing this expression with Eq.~\eqref{eq:S_eta} with $t=N$ we
see that this is the same exponent $\eta$ as was defined in the
diffusion problem for the same type of surface.

Our goal is to establish numerical value of the exponent $\eta$
in a variety of geometries for ideal polymers. In simple geometries
[such as circular cones (in $d=3$) or wedges (in $d=2$)] this
exponent is known analytically
\cite{Ben-Naim2010,Maghrebi2011,Maghrebi2012,Hammer2014}.
(In some cases, $\eta$ is known even for polymers
in good solvents \cite{remark}, where it is found by studying numerically SAWs
in $d=3$ for flat surfaces \cite{Binder83,debell,gsurface} and
for circular cones or cone-plane geometries \cite{Maghrebi2011,Maghrebi2012}.)

Polymers can mediate forces between two surfaces. Consider a polymer
attached by one end to a special point of one SF surface that is brought
close (distance $h$) from a special point of another SF surface. As an
example, we may consider a polymer attached to an apex of a cone that is
being brought into the vicinity of a plane. This resembles measurements done
by means of an atomic force microscope (AFM) \cite{Sarid1994,Giessibl2003},
where a long molecule is attached to the sharp tip of a probe and the
probe is brought into the vicinity of a flat stage. (However, in a typical
experimental situation the {\em other} end of the polymer is also attached
to the stage.) The loss of polymer entropy leads to entropic force between
the two objects \cite{Maghrebi2011,Maghrebi2012}:
\begin{equation}
F=\mathcal{A}\frac{k_{\mathrm B} T}{h},
\label{eq:F}
\end{equation}
where $\mathcal{A}$ is a dimensionless prefactor. This relation is true only
for $h$ much smaller than the root mean squared end-to-end distance of the
polymer. (In the absence of any additional length scales, this is the only
possible dimensionally correct expression for the force.) The change in free
energy of the polymer when bringing one object into contact with the second
one is due to the change in entropy $S=k_{\mathrm B}\ln{\mathcal{N}}$. Both
the initial and final states of the system (but not the intermediate ones)
are SF and can be characterized using the exponents $\eta$. The difference
in the entropy (and free energy) of the two states must be equal to the work
performed by the force. Since the leading exponential part of the number
of states $z^N$, which corresponds to the leading {\it extensive} part of the
entropy and free energy, is identical in both SF situations, the difference
in the free energies is proportional to the difference of the subleading terms $(\gamma_\text{initial}-\gamma_\text{final})\ln N$. Comparison of this relation
with the integral of the force in Eq.~\eqref{eq:F} over the separation $h$
between a microscopic distance $a$ and the maximal interaction distance
$\sim aN^\nu$  fixes the value of the prefactor \cite{Maghrebi2011,Maghrebi2012}:
\begin{equation}
\mathcal{A}=\eta_\text{final}-\eta_\text{initial}.
\label{eq:A}
\end{equation}
Thus, the force between two surfaces, mediated by the polymer, depends
solely on the geometry of the surfaces through the exponents $\eta$
corresponding to the initial and the final states. Equations \eqref{eq:F}
and \eqref{eq:A} have been derived \cite{Maghrebi2011,Maghrebi2012}
by assuming equilibrium conditions. Long equilibration times of
polymers \cite{Gennes1979} make the equilibrium measurements of
force-position relations a difficult task. For {\it slightly}
non-equilibrium energy-dominated experimental situations Crooks
fluctuation theorem \cite{Crooks99} can be used to recover some
equilibrium properties  from non-equilibrium experiments
\cite{Collin05}.

\section{The method}\label{sec:method}
Since calculation of the force constant has been reduced to finding
$\eta$ in SF geometries, our aim is to consider simple methods for
accomplishing this task. For ideal polymers in high symmetry systems,
such as three-dimensional problem of a circular cone
[Fig.~\ref{fig:scalefree}(a)], including its particular cases of plane
or semi-infinite line, or a circular cone attached to a plane
perpendicular to its axis [Fig.~\ref{fig:scalefree}(d)]
it is possible to solve Eq.~\eqref{eq:eta} analytically
(see, e.g., Refs.~\cite{Ben-Naim2010,Maghrebi2011,Maghrebi2012}).
(Such solutions can also be obtained in arbitrary $d$.)
However, even in simple figures such as cones with square or star-shaped
cross section, the solution of the eigenvalue equation becomes
rather cumbersome. Fortunately, there is another, rather simple
numerical approach to the problem described in the following paragraphs.

\begin{figure}[t]
	\centerline{\includegraphics[width=6.5cm]{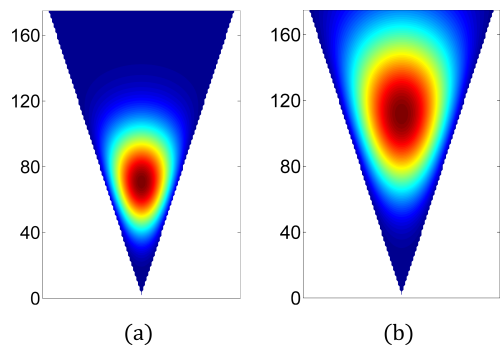}}
\caption{(Color online) Vertical cross section of $P(\vec{r},t)$
of particle diffusing inside circular cone with apex angle
$\alpha=\pi/10$ and absorbing boundaries. At $t=0$ the particle is
located a few lattice constants away from the apex.
Probability density is shown at (a) $t_1=2000$ and (b) $t_2=5000$.
The {\em shape} of the function does not change with time:  by expanding
(a) by a factor $\sqrt{t_2/t_1}$ it will coincide with (b).}
\label{fig:selfsimilar}
\end{figure}

Consider solving the time-dependent Eq.~\eqref{eq:diffeq},
beginning with the $\delta(\vec{r}-\vec{r}_0)$ initial condition for $P$
and evolving the equation in time. At very short times $P$ will resemble
free space solution uninfluenced by the boundaries. If the starting point
is at some distance $r_0$ from an apex of the cone, then
at time $t\sim\tau\equiv r_0^2/D$ the influence of the boundaries
will be strongly felt, and for $t\gg\tau$ the initial conditions will be
forgotten. The solution of the problem will approach \cite{Hammer2014}
\begin{equation}
P(\vec{r},t)=At^{-\eta-d/2}r^\eta
\exp{\left\{-\frac{r^2}{4Dt}\right\}}g(\{\theta_i\}),
\label{eq:prob}
\end{equation}
where only the prefactor $A$ will depend on $\vec{r}_0$. It is possible
to verify directly that this expression solves the differential
Eq.~\eqref{eq:diffeq} {\em exactly}, provided $g$ and $\eta$ are
solutions of Eq.~\eqref{eq:eta}. This solution, however, does not satisfy
the exact initial conditions, and can be used only for sufficiently
long times. In the absence of geometric length scale, the distance
$r$ from the apex in Eq.~\eqref{eq:prob} can be only compared with $\sqrt{Dt}$.
Thus, the shape of the function does not change over time except for being
stretched. Figure~\ref{fig:selfsimilar} depicts the numerical solution
for a diffusing particle inside a circular cone with absorbing
boundaries calculated on a cubic lattice. The simulation begins with
the particle situated on a lattice site close to the apex of the
cone and the probability is then evolved by using discrete diffusion
equation, i.e., the probability at a particular lattice site (inside the
permitted space) at time $t+1$ is equal to the mean of the probabilities
at the neighboring sites at time $t$. The absorbing boundary conditions
are implemented by keeping probability 0 outside the permitted space.
Due to the absorption, the total survival probability decreases. The
color-coding in this and other pictures depicting the density was
chosen to be a linear scale ranging from dark blue (=0) to dark red
(= maximum), thus removing the effect of overall decrease of the function.
Figures \ref{fig:selfsimilar}(a) and \ref{fig:selfsimilar}(b) depict the probability at two
different times. Both functions seem to have the same shape spread
over distances much larger than the few lattice constants that the
initial position $\vec{r}_0$ was separated from the apex of the cone.
This persistence of the shape confirms the claim that the position
of the starting point has been ``forgotten."

\begin{figure}[t]
\centerline{\includegraphics[width=8.5cm]{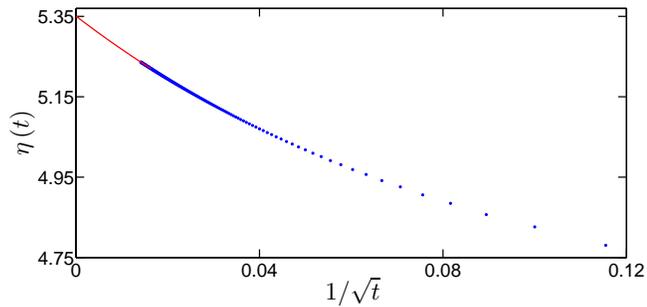}}
\caption{(Color online) Dependence of the effective exponent $\eta(t)$
on simulation time $t$ up to $t=5000$ for a cone with
star-shaped cross-section  [see Fig.~\ref{fig:scalefree}(b)] with
inner apex angle $\alpha=\pi/4$, and the apex angle of the outer point
equal $\pi/\sqrt{2}$. The residual $t$-dependence is evident. By fitting
the data by quadratic polynomial in $1/\sqrt{t}$ (solid line) we
obtain the extrapolated value  $\eta=5.35$.}

	\label{fig:extrapolate}
\end{figure}

As expected, the integral of Eq.~\eqref{eq:prob} over the space produces
$S\propto t^{-\eta/2}$, which can be used to determine the exponent
$\eta$ \cite{Ben-Naim10,Ben-Naim10a}. Numerically, this can be
accomplished by simply summing up all the probabilities at several
times $t$ and extrapolating
the {\em slope} of the graph (on a logarithmic scale) to large
times. The method presented below offers a slightly more convenient
alternative into measuring $\eta$.

We notice that the exponential term in Eq.~\eqref{eq:prob} is exactly
the same as it would be in the absence of boundaries.
The absorbing boundaries generate the time-dependent (power law)
prefactor, as well as term $r^\eta$. While the probability density
in Eq.~\eqref{eq:prob} is not normalized due to absorbtion, we may
calculate the mean squared distance $R^2$ of a {\em surviving}
particle from the apex $R^2=\int r^2P(\vec{r},t){\rm d}^dr/
\int P(\vec{r},t){\rm d}^dr$. In this ratio of integrals, the prefactors
and the angular integrals cancel out, and we are left with the ratio
of two simple integrals leading to
\begin{equation}
R^2=2Dt(\eta+d).
\label{eq:reta}
\end{equation}
Note, that in the absence of boundaries $\eta=0$, and the above result
reduces to the well-known free-space expression $R^2=2dDt$. For large
time, $R^2$ coincides with mean-squared traveled distance by the walker,
or mean squared end-to-end distance of an ideal $N$-step polymer for
$N=t$. Since the numerical evolution of the diffusion equation is
a very simple task, we can use Eq.~\eqref{eq:reta} to calculate
exponent $\eta$, by finding the large $t$ limit of $(R^2/2Dt)-d$.
Of course, for finite times we expect corrections due to discreteness
of the lattice, presence of ragged boundaries on a lattice, and due
to the fact that the walker begins its path few lattice constants
away from the apex. Each of these problems has a typical length scale
of the order of lattice constant $a$, and their influence will
disappear when the dimensionless ratio $a/R\sim1/\sqrt{t}\ll 1$.
Naturally, we expect the $t$-dependent effective exponent
$\eta(t)=\eta+c_1/\sqrt{t}+c_2/t+\cdots$, where the first term
is the anticipated actual value of the exponent.
Figure \ref{fig:extrapolate} depicts the time-dependence of $\eta(t)$
for a cone of star-shaped cross-section measured up to $t=5000$.
Last calculated point is only few percent away from the intercept
(``$t\to\infty$'') with the vertical axis at  $\eta=5.35$, which was
obtained by  extrapolating the numerical values using a quadratic
polynomial in $1/\sqrt{t}$. Our method requires solution of the
diffusion equation, i.e., its numerical complexity does not differ
from the calculation relying on the measurement of the power-law
dependence of the survival probability mentioned in the previous
paragraph. However, we believe that it provides a slightly more
convenient alternative.

\begin{figure}[t]
\centerline{\includegraphics[width=8.5cm]{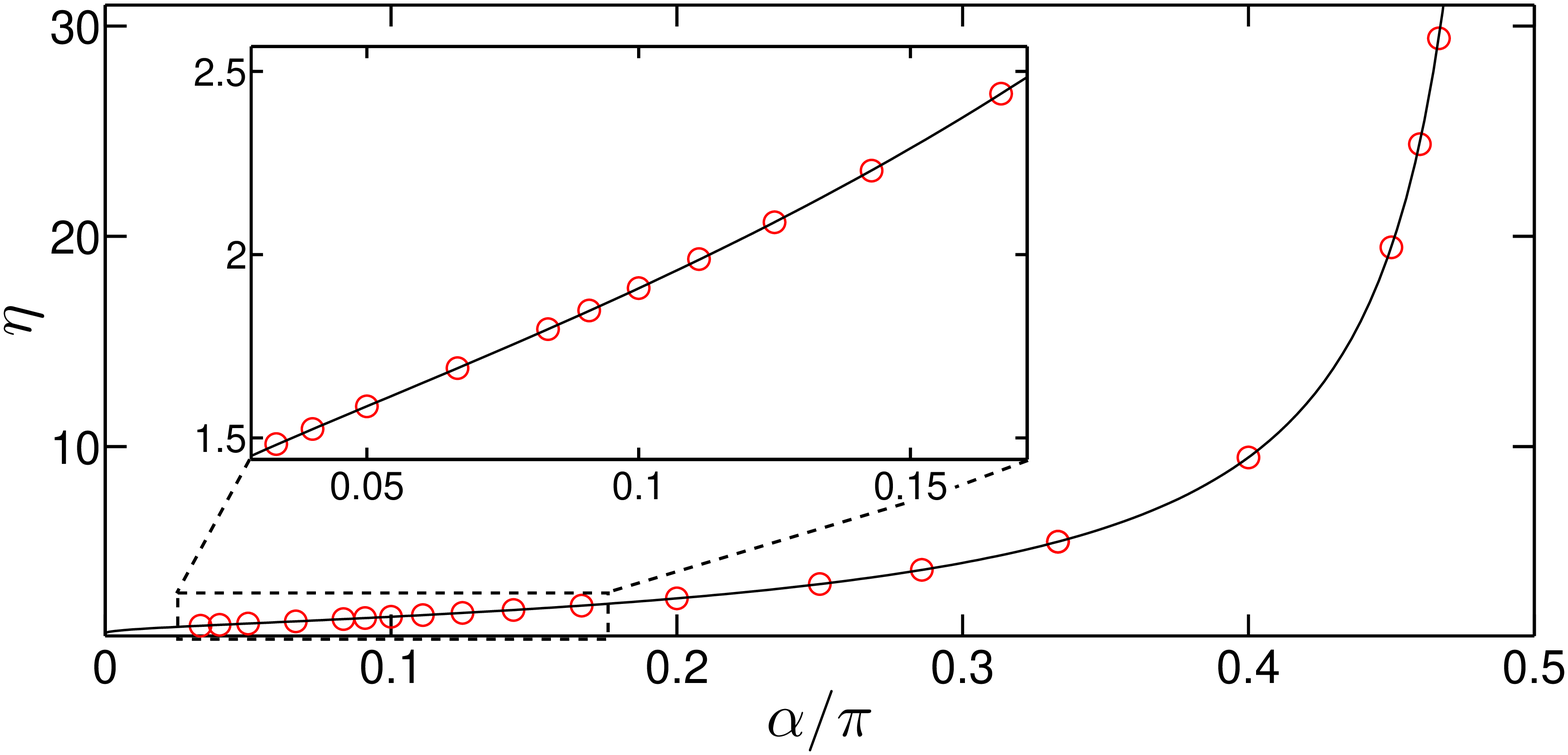}}
\caption{(Color online) Exponent $\eta$ as function of apex angle
$\alpha$ for cone-plane geometry depicted in Fig.~\ref{fig:scalefree}(d).
Our numerical results (circles) are compared with the known
analytical values \cite{Maghrebi2012} (solid line).}
	\label{fig:CPgraph}
\end{figure}

Exponent $\eta$ for an ideal polymer attached to a contact point
between a plane and a cone perpendicular to that plane, as depicted
in Fig.~\ref{fig:scalefree}(d), can be found analytically
\cite{Maghrebi2011,Maghrebi2012}. In this three-dimensional geometry
with azimuthal symmetry the eigenvalue Eq.~\eqref{eq:eta} becomes
Legendre equation, and the eigenfunction $g(\theta)$ can be expressed
as a linear combination of regular Legendre functions of degree $\eta$.
The value of $\eta$ in this cone-plane geometry depends on the apex angle
$\alpha$ of the cone and is determined as the (smallest) value for which
the absorbing boundary condition ($g(\theta=\alpha)=0$) is satisfied
\cite{Maghrebi2012}. Figure \ref{fig:CPgraph} depicts the results of numerical
evaluation of the exponent compared with the known analytical values.
Excellent correspondence of the results validates our numerical procedure.
As the apex angle increases, the number of polymer configurations trapped
between the cone and the plane that it touches decreases as can be seen
in the increasing value of $\eta$. For $\alpha\to\pi/2$ the exponent
$\eta$ diverges, as expected. As $\alpha$ increases, so does the numerical
difficulty to obtain accurate estimate of $\eta$, requiring larger
times $t$.

\begin{figure}[t]
\centerline{\includegraphics[width=8.5cm]{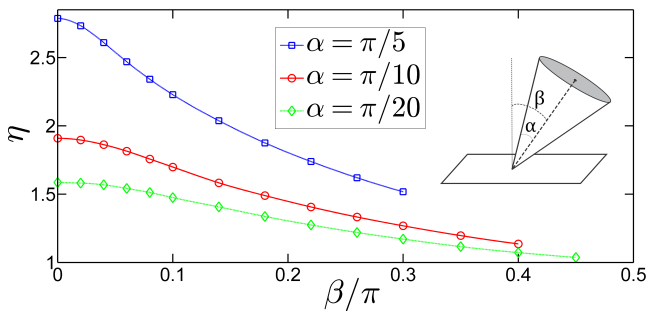}}
\caption{(Color online) Exponent $\eta$ as function of tilt angle $\beta$
for a tilted circular cone touching a plane for three apex angles
(top to bottom) $\alpha=\pi/5$ (square), $\alpha=\pi/10$ (circle), and
$\alpha=\pi/20$ (diamond). The solid lines interpolate between numerical
results.}
\label{fig:TCPgraph}
\end{figure}

\section{Exponent $\eta$ for geometries without azimuthal symmetry}\label{sec:results}

\begin{figure}[t]
\centerline{\includegraphics[width=8.5cm]{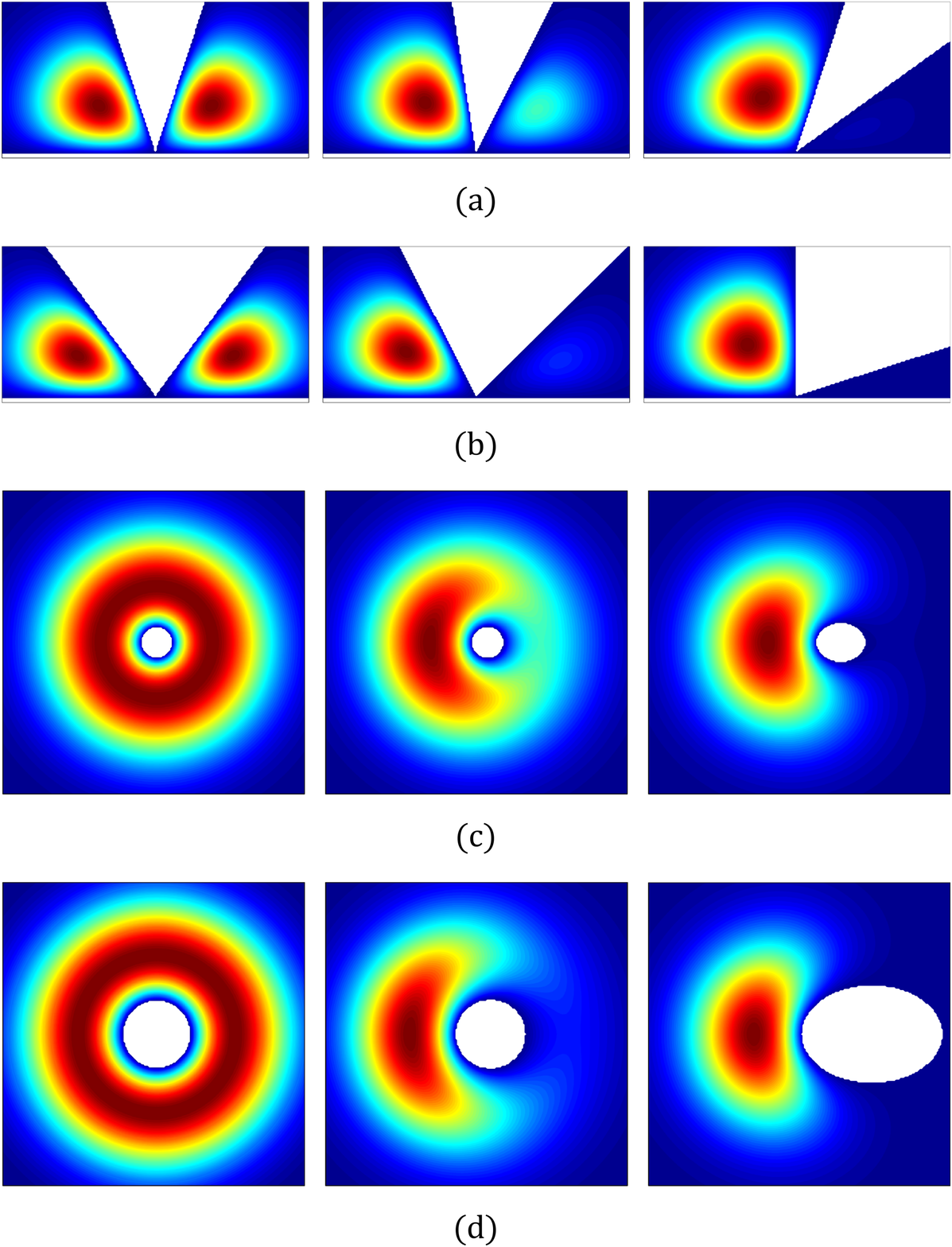}}
\caption{(Color online) Probability density $P$ for diffusing particle in
the vicinity of cone touching a plane.  Cross-sections perpendicular to the
plane contacted by the cone include the axes of the cones.  Two
apex angles (a) $\alpha=\pi/10$ and (b)
$\pi/5$ are considered. Panels (c) and (d) show
cross-sections parallel to the plane contacted by the cone for those
apex angles. This section is made at height 30 lattice constants at time
$t=4000$. This is the level where $P$ is close to its maximum value. The
left column shows  upright cone ($\beta=0$), while the center and right
columns show tilts of $\beta =0.05\pi$ and $0.15\pi$, respectively.
}
\label{fig:TCPCS}
\end{figure}

While analytical values of $\eta$ could be found for the cone perpendicular
to the plane considered in the previous section,
tilting the cone axis by angle $\beta$ with respect to the normal to
the plane (see sketch in Fig.~\ref{fig:TCPgraph}) breaks the azimuthal
symmetry, and prevents a simple analytical solution. We therefore find
values of $\eta$ as function of $\beta$ numerically, for $\alpha=\pi/5$,
$\pi/10$, and $\pi/20$, as depicted in Fig.~\ref{fig:TCPgraph}. The tilts
are, of course, limited by the inequality $\beta\le(\pi/2)-\alpha$. The maximal
value of $\eta$ for any $\alpha$ is achieved for an upright cone, and the
values decrease with increasing tilting. For a polymer attached to
a planar surface, i.e., in the case of completely absent cone, $\eta=1$.
The cone with $\alpha=\pi/20$ almost reaches that value
when it is tilted to the maximal extent. The other two cases also exhibit
a significant  decrease with increasing $\beta$. This means that an upright
cone, causes the maximal constriction of the space available to the polymer,
while the tilt significantly decreases that effect. Figure \ref{fig:TCPCS}
depicts the cross-sections probability density in planes parallel and
perpendicular to the plane that is touched by the cone for two apex
angles $\alpha$. We can clearly see that even for small tilts the
polymer ``escapes" to the more open part of the space, where its
configurations somewhat resemble the behavior of a polymer in half-space
in the absence of a cone.

\begin{figure}[t]
\centerline{\includegraphics[width=8.5cm]{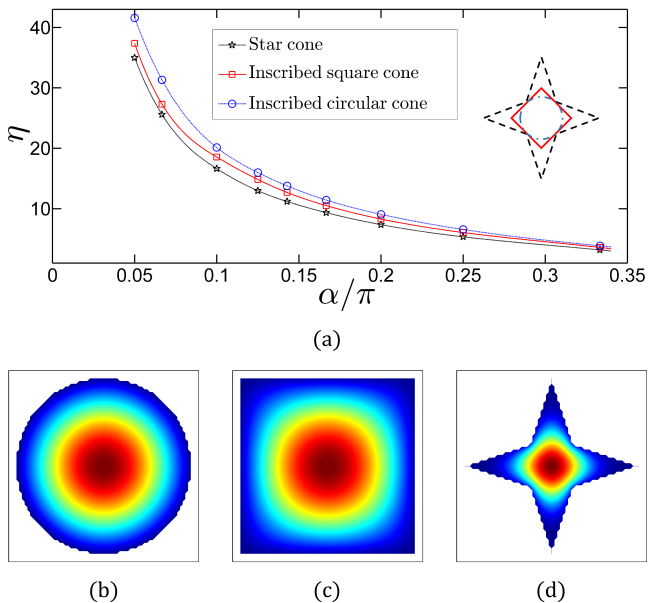}}
	\caption{(Color online) (a) Exponent $\eta$ as function of
apex angle $\alpha$ of a circular cone and for cones of square- and
star-shaped cross-section circumscribed around the circular cone as
depicted  in Figs.~\ref{fig:scalefree}(a)-\ref{fig:scalefree}(c), as
well as in the embedded sketch. The outer radius of the star-shaped
cone is $2^{3/2}$ times larger than its inner radius. Symbols indicate
the numerical values, while the continuous lines interpolate
between the calculated values.  Bottom part of the figure depicts
transverse cross-sections of $P(\vec{r},t)$ for inside
(b) circular, (c) square-shaped, and
(d) star-shaped cones. All sections were performed 100
lattice constants away from the apex for $t=4000$ and
$\alpha$ of a circular cone equal to $\pi/10$.}
\label{fig:cones}
\end{figure}

Finally, we consider diffusing particles inside cones with circular, square,
and four-point star cross-sections
[Figs.~\ref{fig:scalefree}(a)-\ref{fig:scalefree}(c)]. Figure
\ref{fig:cones}(a) depicts the dependence of the exponent $\eta$
for several cross-section shapes as a function of an opening angle,
while Figs.~\ref{fig:cones}(b)-\ref{fig:cones}(d) show transverse
cross sections of $P(\vec{r},t)$ for the three different shapes.
An interesting feature is seen in Fig.~\ref{fig:cones}(d), where $P(\vec{r},t)$
inside the star-shaped cone resembles that of diffusion
inside the square-shaped cone, rotated by $45^\circ$. This implies that the
surviving(!) diffusing particle is less likely to be found deep inside the
star's wings and can be found in the center of the star with much higher
probability. This feature
is also seen in Fig.~\ref{fig:cones}(a), where the values of $\eta$s are
similar, though not identical, in the two geometries.

\section{Discussion}
In this paper we examined the long-time solutions of the diffusion
equation in the presence of SF absorbing boundaries and found the
probability density function of a diffusing particle. Unlike finite
spaces that are characterized by finite absorption times
and exponential decay of survival probability,  SF absorbing shapes
generate survival functions that decay as power law $t^{-\eta/2}$.
The same exponent appears in the spatial part of the probability
density of a diffusing particle, and can be calculated from the
measurement of mean-squared distance traveled by surviving
particle. Our results provide information regarding the behavior of
ideal polymers. In high space dimension the solid angle describing
the cone plays a major role in determining the value of $\eta$, while
specificities of the shape are not very important
\cite{Ben-Naim10a,Ben-Naim14}. Our results with tilted cones, as seen
in Fig.~\ref{fig:TCPgraph}, or with cones with various cross-sections
demonstrate the sensitivity of the values of $\eta$ on the shape
details in $d=3$ for a fixed solid angle.

In real experiments we are more likely to encounter
polymers described by different statistics. In particular, polymers
in good solvents are better described by self-avoiding walks
\cite{Gennes1979}. For ideal polymers the simple relation
between $R^2$ and the exponent $\eta$ as shown in Eq.~\eqref{eq:reta}
was a specific  consequence of long-time solution in Eq.~\eqref{eq:prob},
which is a product of a power law and a Gaussian with respect to
variable $r$. Polymers in good solvents have a slightly different
functional dependence on $r$, which does not permit a simple
identification of $\eta$ from the probability distribution of their
end-point for a single value of $N$. Nevertheless, the general
expression for the force constant in Eqs.~\eqref{eq:F} and \eqref{eq:A}
remains valid, while the values of the exponents $\eta$ maintain
similar shape dependencies and are surprisingly {\em close numerically}
to the values of $\eta$s for ideal polymers
\cite{Bubis2009,Maghrebi2011,Maghrebi2012}. Therefore,
our results, besides providing some intuition regarding the polymers
in good solvents, also provide reasonable guesses for the numerical
values of the exponents in such solvents.

\begin{acknowledgments}
We thank Y. Hammer and M. Kardar for numerous discussions.
This work was supported by the Israel Science Foundation Grant No.~186/13.
\end{acknowledgments}

\bibliographystyle{apsrev}

\begin{thebibliography}{31}
\bibitem{remark}
 While our paper deals exclusively with ideal polymers, similar scaling forms (albeit with different exponents) apply to real polymers or SAWs. Indeed, preliminary work indicates striking qualitative (and even quantitative) similarities for SAWs.
\end{thebibliography}

\end{document}